\documentclass[reprint,bibnotes,amsmath,amssymb,aps,prb,showpacs,floatfix,superscriptaddress,longbibliography]{revtex4-1}
\usepackage[breaklinks=true,colorlinks,citecolor=blue,linkcolor=blue,urlcolor=blue]{hyperref}
\usepackage{epsfig,mathrsfs,color,latexsym,subfigure,marginnote}

\renewcommand{\BibitemShut}[1]{}

\newcommand{\nn}{\nonumber}

\newcommand{\ct}{\cite}

\newcommand{\be}{\begin{equation}}
\newcommand{\ee}{\end{equation}}
\newcommand{\ba}{\begin{eqnarray}}
\newcommand{\ea}{\end{eqnarray}}

\begin{document}
\title{Dynamics of edge current in a linearly quenched Haldane model}
\author{Sougata Mardanya}
\affiliation{Department of Physics, Indian Institute of Technology, Kanpur, Kanpur 208016, India}
\author{Utso Bhattacharya}
\affiliation{Department of Physics, Indian Institute of Technology, Kanpur, Kanpur 208016, India}
\author{Amit Agarwal}
\email{amitag@iitk.ac.in}
\affiliation{Department of Physics, Indian Institute of Technology, Kanpur, Kanpur 208016, India}
\author{Amit Dutta}
\email{dutta@iitk.ac.in}
\affiliation{Department of Physics, Indian Institute of Technology, Kanpur, Kanpur 208016, India}

\begin{abstract}
In a finite time quantum quench of the Haldane model, the Chern number determining the topology of the bulk remains invariant, as long as the dynamics is
unitary. Nonetheless, 
the corresponding boundary attribute, the edge current, displays interesting dynamics. For the case of sudden and adiabatic quenches the post quench edge current is solely determined by the initial and 
the final Hamiltonians, respectively. However for a finite time ($\tau$) linear quench in a Haldane nano ribbon, we show that the evolution of the edge current from the sudden to the adiabatic limit is not monotonic in $\tau$, and has a turning point at a characteristic time scale $\tau=\tau_c$. 
For small $\tau$, the excited states lead to a huge unidirectional surge in the edge current of both the edges. On the other hand, in the limit of large $\tau$, the edge current saturates to
its expected equilibrium ground state value. This competition between the two limits lead to the observed non-monotonic behavior. Interestingly, $\tau_c$  seems to depend only on the Semenoff mass and the Haldane flux. A similar dynamics for the edge current is also expected in 
other systems with topological phases.
\end{abstract}
\maketitle

\section{Introduction} 


Subtle topological phenomena such as 
the imaging of the edge states in  
cold atomic quantum Hall systems \ct{stuhl15}, the direct measurement of the
Berry curvature \ct{jotzu14} and the Zak phase \ct{atala13} have been demonstrated in cold atomic topological bands. 
Beyond these static situations, the dynamical topological properties of systems following a quantum quench 
have also been experimentally probed \ct{killi12,hauke14,grushin16}. A quantum quench, forces the initial state prepared in the 
quantum many body ground state of the initial Hamiltonian, to undergo dynamical evolution far from equilibrium \ct{greiner02}. Thus 
quantum quenches offer the promise of engineering distinct many-body non-equilibrium states which have 
no equilibrium counterpart \ct{oka09,kitagawa11,rudner13,torres14,dehghani15,wilson16}. This has motivated a plethora of 
studies of the non-equilibrium
dynamics of both closed and open topological quantum systems under the
application of quantum quenches \ct{sharma12,heyl13,karrasch13,budich16,bhattacharya17a,heyl15,palmai15,vajna14,sharma15,vajna15,
schmitt15,sharma16,divakaran16,bhattacharya17b,dutta17,mukherjee12,suzuki16,dora13,
sachdeva14,bhattacharya17c} and periodic drives \ct{thakurathi13,thakurathi14,mukherjee09,das10,quan06,nag12,sharma14,nag14,
agarwala16,lazarides15,bhattacharya16,dasgupta15,bhattacharya17d}.

More recently there has been a  significant interest in the dynamics of the edge current following a quantum quench in 
a system taking it either from a topological phase to a trivial insulator phase or the vice versa. 
Studying the Haldane model a  hexagonal lattice \ct{haldane88}, Ciao {\it et.~al.} \ct{caio15}, showed that the Chern number 
of the initial phase in the translationally invariant Haldane model remains preserved throughout the post quench unitary evolution of the system, 
irrespective of the topology of the final Hamiltonian.
The invariance of the Chern number under any unitary dynamics, has also been rigorously established by Alessio {\it et.~al.}\ct{alessio15}. The invariance of the Chern number has also been shown 
for the quantum quench in the Haldane model with higher order hoppings \ct{sticlet13,bhattacharya17e}. 
The preservation of the winding number of the many-body state was also mentioned \cite{foster14, foster13} in the context of  quenches in interacting topological BCS superfluids. However,  the dynamics of the edge current following quench in these systems and the fate of the corresponding `bulk-boundary correspondence' 
still remains an interesting open question. 
Motivated by this, in this article we study the dynamics of the edge current in \textit{a finite time linear quench} across the topological phase transition point, 
by varying the Semenoff mass in a Haldane nano-ribbon.

Earlier studies on the Haldane model \ct{caio15}, and the Haldane model with higher hopping \ct{sticlet13}, showed that following a global \textit{sudden} quench from a 
topological to non-topological phase, the edge current relaxes from a finite value to a post quench value close to zero - which is the 
value corresponding to the ground state of the final Hamiltonian. In this article we focus on the role  of the finite rate  of the quench on the dynamics of the edge current, 
by considering a linear finite time quenching of the Semenoff mass \cite{PhysRevLett.53.2449} in the Haldane nano-ribbon \ct{haldane88}, taking it from a topological to trivial phase. 
Interestingly, we show that in quenching from non-topological phase to the topological phase, the edge current evolves in a non-monotonic way as a function 
of the quenching rate ($\tau$) and has a turning point on increasing  $\tau$ from the sudden ($\tau=0$) to the adiabatic ($\tau \to \infty$) limit. 

The paper is organized as follows: The equilibrium Haldane model in a translationally invariant system  in both the directions is introduced in Sec.~\ref{section2}; It also discusses the
equilibrium  edge current for the model on a  nano-ribbon geometry, periodically wrappped in the $x$-direction and open in the $y$-direction. This is followed by a discussion of the impact of a sudden and adiabatic quench, driving the system from non-topological phase to topological phase, on the edge current in Sec.~\ref{section3}. The role of a finite time linear quench on the dynamics of the edge current is described in Sec.~\ref{S4}.
Finally we summarize our findings in section \ref{section4}. 

\section{The Haldane Model}\label{section2}
\begin{figure}[t]
\includegraphics[width=0.90 \linewidth]{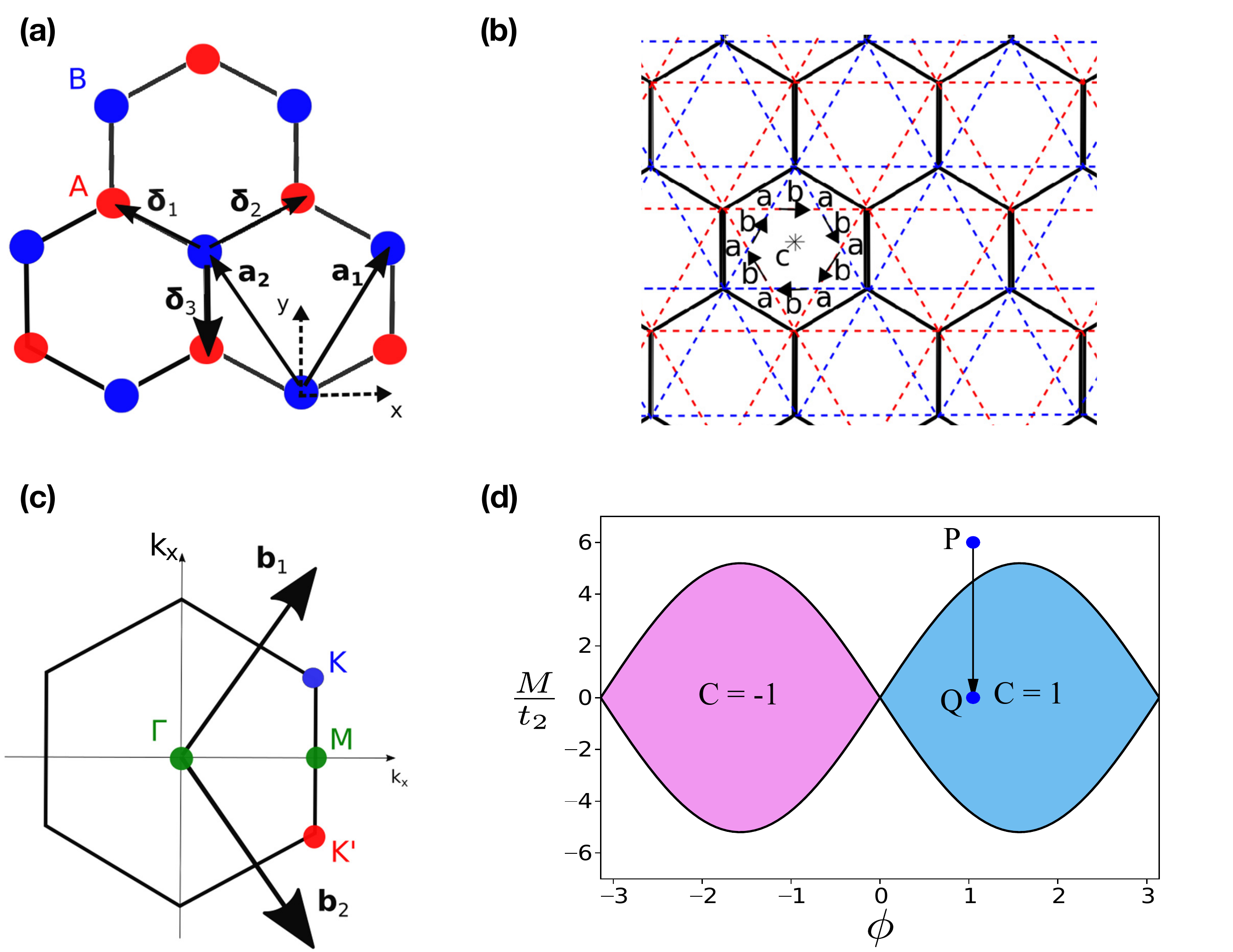}
\caption{(a) Three plaquettes of the hexagonal lattice with lattice vectors ${\bf a_1}$ and ${\bf a_2}$. The blue and red dot represent the two sub-lattices A and B. (b) The Haldane lattice, with locally broken time reversal symmetry. 
 The regions marked by $a$ and $b$ enclose flux in opposite directions, and the arrows on the next nearest bonds 
(dashed lines) denote the direction of the positive phase hopping due to the locally broken time reversal invariance. 
(c) The BZ of hexagonal lattice with reciprocal lattice vector ${\bf b_1}$ and ${\bf b_2}$ with ${\bf K}$ and ${\bf K}^{\prime}$ representing the two inequivalent Dirac points, with the color corresponding to the sub-lattices A and B. 
(d) The Chern phase diagram of the Haldane model in the on-site energy $M$ (also called the Semenoff mass) and the staggered phase $\phi$, plane. The white region is the topologically trivial phase ($\nu = 0 $), while the colored region is the 
topologically non trivial Chern phase ($\nu = \pm1$). The arrow indicates a quenching scheme of varying $M$, which takes the model from the point $P$ ($\phi=\pi/3$,$M/t_2=6$) in the non-topological phase  to the point Q ($\phi=\pi/3$,$M/t_2=0$) in the topological phase.
}\label{fig1}
\end{figure}

Our starting point is the Haldane model with broken spatial inversion and locally broken time reversal symmetry, describing the nearest and next nearest hopping of spin-less electrons on a Hexagonal lattice. The 2D hexagonal graphene-like lattice composed of the two triangular sub-lattices A and B is shown in Fig. \ref{fig1}(a). 
The Hamiltonian of this model is explicitly given by,
\begin{eqnarray}\nonumber
\label{eq_ham}
H=& &\sum_{\langle i,j \rangle}t_1 \left(c_{iA}^{\dagger}c_{jB} + h.c\right) + M \sum_{i\in A}\hat{n}_{i}-M\sum_{i\in B}\hat{n}_{i}\\ 
&+& \sum_{\langle\langle i,j \rangle\rangle}t_2 e^{i\phi_{ij}} \left(c_{iA}^{\dagger}c_{jA}+c_{iB}^{\dagger}c_{jB} + h.c\right)~,
\end{eqnarray} 
where $c^{\dagger}_i$ ($c_i$) is the fermionic creation (annihilation) operator at site $i$ satisfying the anti-commutation relation $\{c_i^{\dagger}, c_j\}=\delta_{ij}$. In Eq.~\eqref{eq_ham}, $\hat{n}_i=c^{\dagger}_i c_i$ and A/B denote the two different sub-lattices. The phase factor 
$\phi_{ij}=\pm \phi$, is positive for anticlockwise hopping and negative for clockwise hopping. It mimics a staggered magnetic field, introduced to break the local time reversal symmetry. Note that the total net magnetic flux through each  hexagonal plaquette is zero, conserving the global time reversal symmetry. This staggered magnetic field, breaking the local time reversal symmetry is what renders the model topologically non-trivial. On the other hand, the different (Semenoff) mass terms $M$ ($-M$) on the two sub-lattices - A (B) - break the spatial inversion symmetry of the model. 
%
%

The real space tight-binding Hamiltonian of Eq.~\eqref{eq_ham}, for a translationally invariant system, i.e., with periodic boundary condition in both the directions, can also be expressed in the crystal-momentum space via a Fourier transform and is given by 
\begin{eqnarray}
& & \mathit{H}= \left(\begin{matrix}
c_{{\bf k}A}^{\dagger} & c_{{\bf k}B}^{\dagger}
\end{matrix}\right) h({\bf k}) \left(\begin{matrix} c_{{\bf k}A} \\ c_{{\bf k}B}
\end{matrix}\right),\label{eq_hamk} \nonumber \\
& & {\rm where,}~~~~ h({\bf k}) = \sum_{i=0}^{3} h_i({\bf k}) \sigma_i~.
\end{eqnarray}
Here $\sigma_i$ (for $i= 1, 2, 3$) are the three Pauli spin matrices, and $\sigma_0$ is the $2 \times 2$ identity matrix. The components  $h_i({\bf k})$ are,
\begin{eqnarray} \nonumber
h_0({\bf k})&=&2t_1\cos \phi \left[{\bf k}\cos({\bf k}.{\bf a_1})+\cos({\bf k}.{\bf a_2}) \right. \nonumber \\
& & \left. +\cos({\bf k}.({\bf a_1}-{\bf a_2}))\right], \nonumber \\
h_1({\bf k})&=& t_1[1+\cos({\bf k}.{\bf a_1})+\cos({\bf k}.{\bf a_2})], \nonumber \\
h_2({\bf k})&=&t_1[\sin({\bf k}.{\bf a_1})+\sin({\bf k}.{\bf a_2})], \nonumber \\
h_3({\bf k})&=&M+M_H,  \\
M_H({\bf k})&=&2t_2\sin\phi[\sin({\bf k}.{\bf a_2})-\sin({\bf k}.{\bf a_2}) \nonumber \\
&&+\sin({\bf k}.({\bf a_1}-{\bf a_2}))]. \nonumber \label{eq_hamt}
\end{eqnarray}
Here $M_H({\bf k})$ is the staggered field and crystal momentum dependent Haldane mass and ${\bf a_1}=\frac{a}{2}(\sqrt{3}, 3)$, ${\bf a_2}=\frac{a}{2}(-\sqrt{3}, 3)$ as shown in Fig.~\ref{fig1}(a).
For $M=0$ and $\phi=0$ the Hamiltonian in Eq.~\ref{eq_ham} reduces to the second nearest neighbour tight-binding Hamiltonian of graphene which has a Dirac like dispersion at six points in the hexagonal Brillouin zone, with only two of them being inequivalent. These two inequivalent points are time reversed partners of each other [see Fig.~\ref{fig1} (b)]. The  other Dirac points are related to these two via reciprocal lattice vectors. 

{Qualitatively, when the local time reversal symmetry breaking term ($\phi$) dominates over the inversion symmetry breaking term ($M$) in the translationally invariant Haldane model, it is topologically characterized by a bulk Chern number which takes the value $\nu = \pm 1$ and zero otherwise}. The Chern phase diagram of the Haldane model is shown in Fig.~\ref{fig1}(d). When the Chern number of the bulk system is $\pm 1$, the boundary of the finite sized open system hosts charge conducting edge modes, consistent with the bulk-boundary correspondence. 
\subsection{Edge Current}
\begin{figure}[t]
\includegraphics[width=0.8 \linewidth]{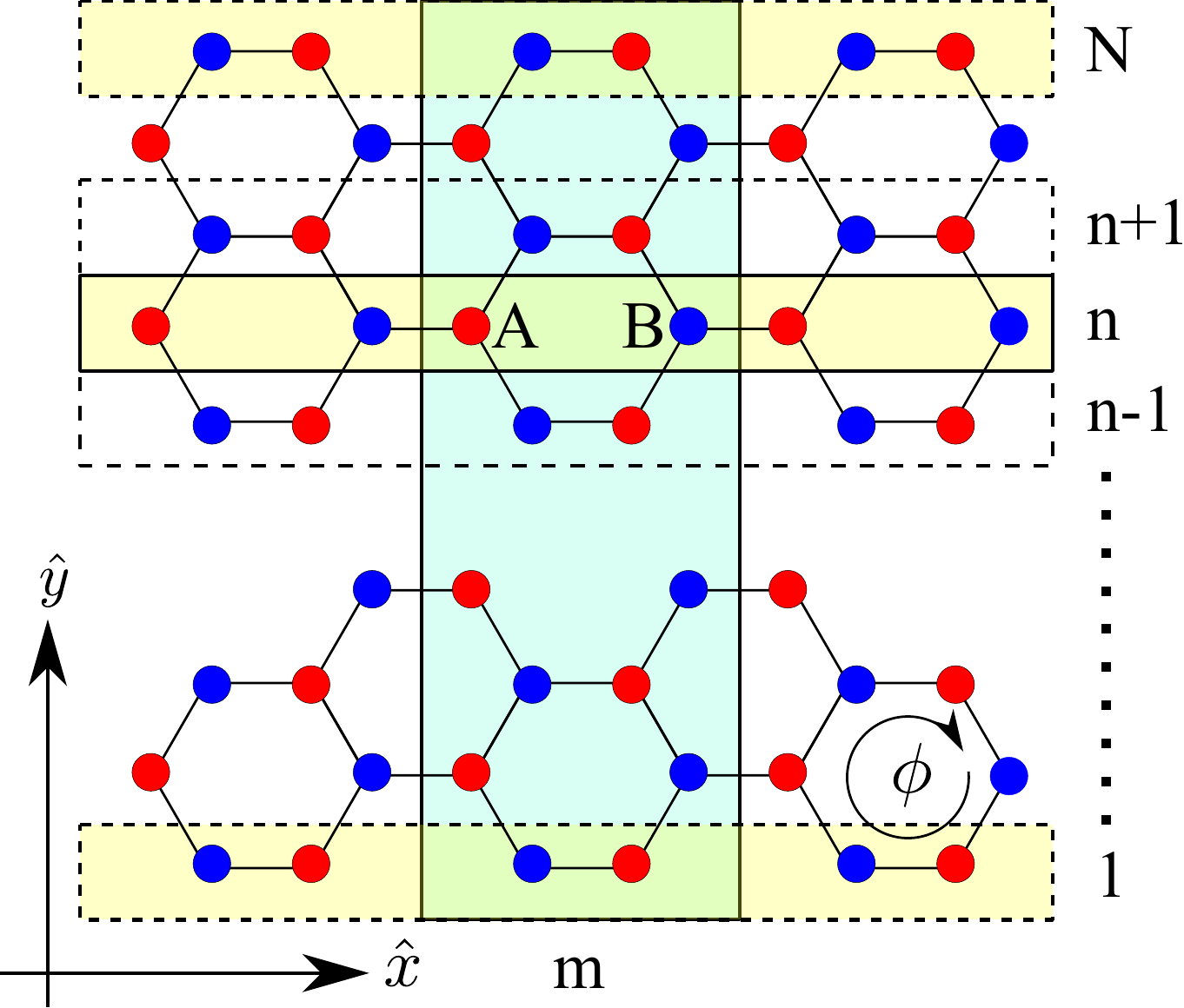}
\caption{The Haldane model on a nano-ribbon, which is periodic along the armchair edge ($x$ direction), and finite along the zigzag direction ($y$ direction). The nano-ribbon is $N$ cells wide and each unit cell is labeled by two indices $m$ and $n$, and has two lattice sites  ($A$ or $B$). 
The phase† of the complex hopping $t_2$ is  negative (positive) for hopping in a (anti) clockwise sense between next-nearest neighbours. 
}\label{fig_lattedge}
\end{figure}

To explore the dynamical evolution of the edge current following a quench, we consider the edge states of the Haldane model which is periodic  (and thus translationally invariant) in the $x-$direction, and has finite width $N$ along the $y-$direction with an armchair edge. 
The schematic for the same is depicted in Fig.~\ref{fig_lattedge}.  Using the conserved crystal momentum along the periodic $x-$direction, while retaining the real space description in the $y-$direction, we obtain the following Hamiltonian 
\begin{eqnarray}
\label{eq_hamo}
& & H=\sum_{m,n=1}^{L}~\sum_{m',n'=n-1}^{n+1}~\sum_{s=A,B} e^{i k_x x_{mns}} c^{\dagger}_{k_x n s} \times  \nonumber\\
& &\Big\{M_s c_{k_x n s}+ t_1 c_{k_x n^{\prime} \bar{s}} e^{-i k_x x_{m^{\prime}n^{\prime}{\bar s}}} \nonumber\\
& & +t_2 e^{i\phi_{mm^{\prime}nn^{\prime}}}  c_{k_x n^{\prime}
s} e^{-i k_x x_{m^{\prime}n^{\prime}s}}\Big\},
\end{eqnarray}
where $\bar{s} \neq s$, and $M_s = + M$ ($-M$) for $s=A$ ($s=B$).

\begin{figure}[t]
\includegraphics[width=0.8 \linewidth]{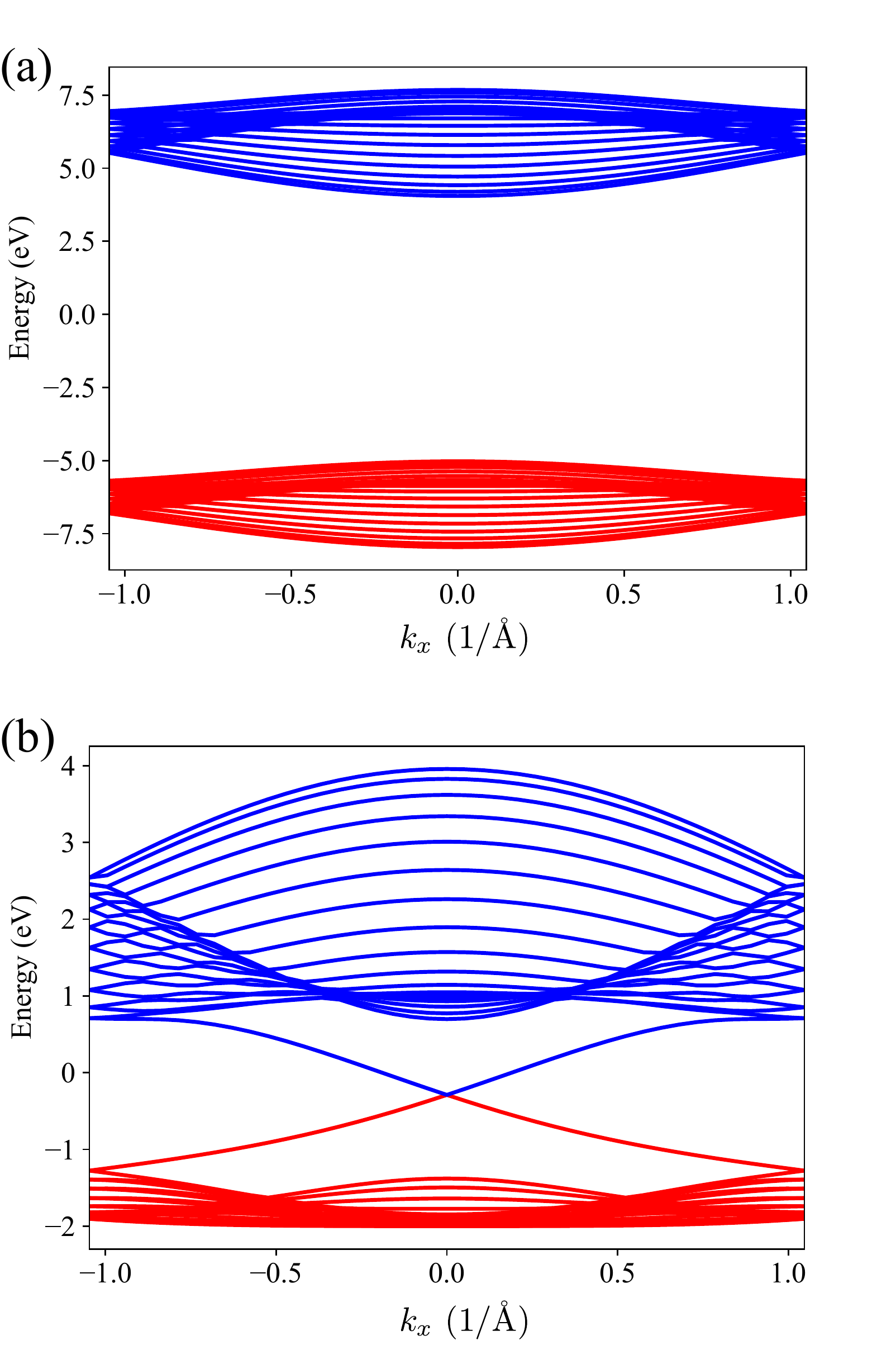}
\caption{The energy spectrum of the Haldane nano-ribbon described by the Hamiltonian in Eq.~\ref{eq_hamo} in (a) the topologically trivial phase with $M = 6$ and the Chern number $\nu =0$, and (b) the topological phase with $M = 0$ and $\nu = - 1$. The blue and red lines represent the conduction and the valance band, respectively. Note that unlike the topologically trivial phase, the topological phase has a single band crossing at $k_x = 0$. We have chosen other parameters to be  $t_1 = 1$, $t_2 = 1/3$, $\phi = \pi/3$, and $N =20$.}
\label{fig_en}
\end{figure}

The energy spectrum of this semi-open model for both the topological and the trivial phase is shown in Fig.~\ref{fig_en}. To clearly exhibit a correspondence between the bulk Chern number (in a periodically wrapped system along both the directions) and the number of mid gap band crossings in the semi-open system, we choose two different sets of parameter values corresponding to the two different phases in the phase diagram in Fig.~\ref{fig1}(d) with $\nu = 0~  \mbox{and} -1$. In the non-topological phase, i.e., for $\nu = 0$ there are no band crossings in the spectrum [see Fig.~\ref{fig_en}(a)], indicating the absence of conducting edge states. On the other hand when we are in the topologically non-trivial phase with Chern number $\nu = -1$, the spectrum in Fig.~\ref{fig_en}(b) clearly shows a mid-gap band crossing between the valance band and the conduction band at $k_x=0$. 

The local current operator at any site $i$ is given by 
\begin{equation}
\hat{J}_i = -\frac{i}{2}\sum_{j}\vec{\delta}_{ij}(t_{ij}{c}_i^{\dagger}{c}_j-h.c.)\label{eq_curr}
\end{equation}
where $t_{ij}$ and $\vec{\delta}_{ij}$ are the hopping amplitude and vector displacement between site $i$ and $j$ respectively. The sum involving the index $j$ is over the nearest and next nearest neighbour sites to $i$ only. Each site of this ribbon is labeled by $\{m,n, s\}$, where $\{m,n\}$ denotes the position of the site in the 2-D lattice and $s$ is the sub lattice index (A or B) of that site. The total current flowing along the strip in the $x$ direction for a particular value of $n$ (where $n$ labels each horizontal row along the $y$ direction -- see Fig.~\ref{fig_lattedge}) is obtained from the following relation,
$$J^x_n = \langle \hat{J}^x_n \rangle = \sum_{k_x,s}{\langle\hat{J}^x_{n,k_x,s}\rangle},$$
where the expectation is taken over the ground state of the Hamiltonian under equilibrium conditions and over the dynamically evolved ground state of the system in case of a quenched system \ct{bhattacharya17e}. 
\begin{figure}[t]
\includegraphics[width=0.9 \linewidth]{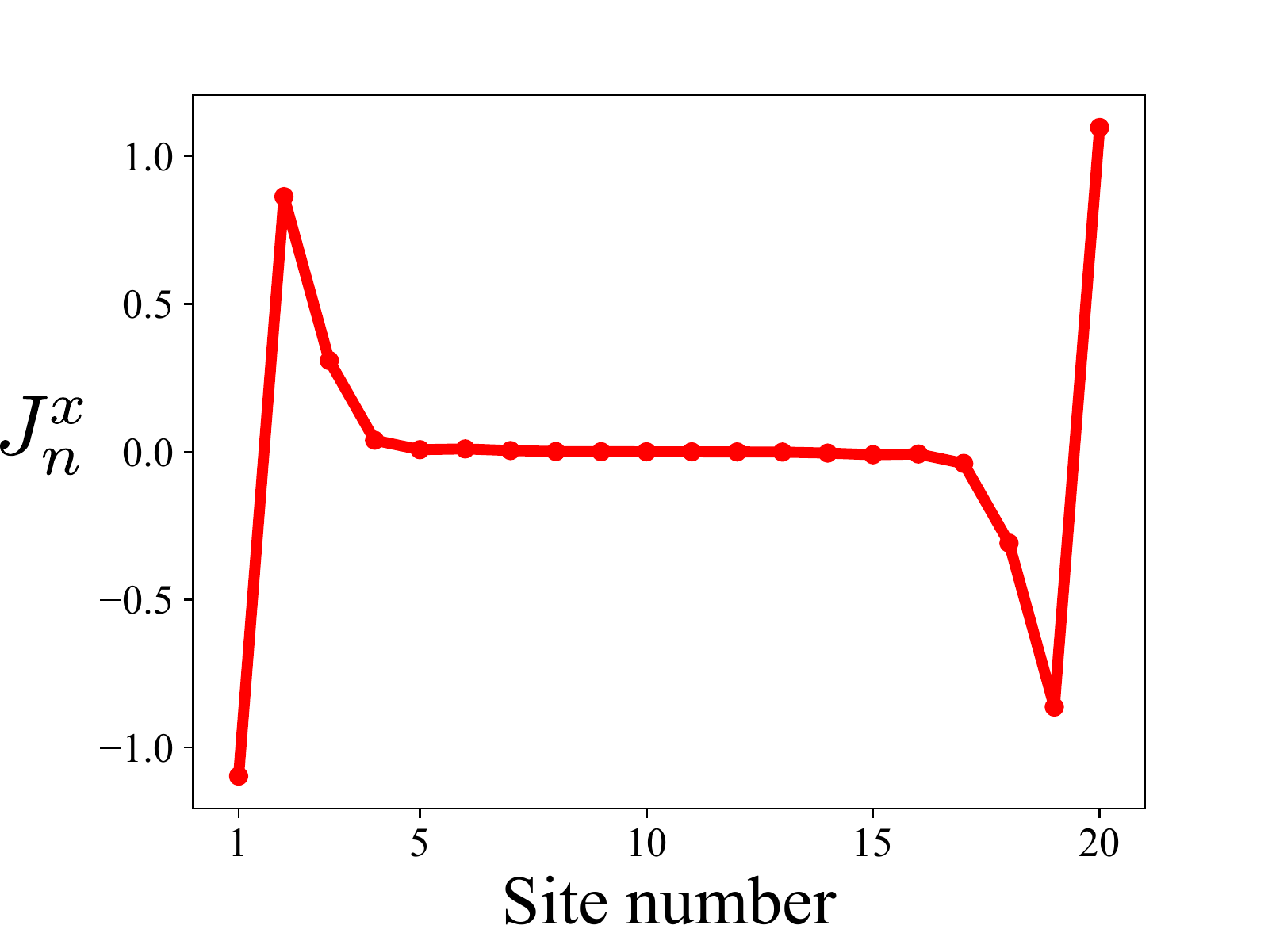}
\caption{The average equilibrium current along the $x$-direction along each row, $n=1,\ldots, 20$ of the Haldane nano-ribbon (see Fig.~\ref{fig_lattedge}) in the topological phase.  Two counter propagating current carrying states appear at the two edges. On the other hand the current remains zero in the bulk. 
Here we have chosen $t_1 = 1$, $t_2 = 1/3$, $M=0$, $\phi = \pi/3$ and $N=20$.}\label{fig_curreq}
\end{figure}

Figure~\ref{fig_curreq} shows the average equilibrium current in the $x-$direction plotted against $n=1,\ldots,N$ when the system is in $\nu = 1$ phase ($M=0$ and $\phi = \pi/3$). As expected, there are two counter propagating channels of current near the system edges (at $n=1$ and $N$) while the current in the bulk is zero throughout. Moreover, the equilibrium current for the topologically trivial gapped phase with $\nu = 0$ is identically zero throughout the system. 

\section{Edge current dynamics following sudden and adiabatic quantum quench}
\label{section3}

In order to investigate the non-equilibrium dynamics of the edge current of the Haldane model, we consider quantum quenches between different points ($M, \phi$) of the phase diagram shown in Fig.~\ref{fig1}(d). The  values of the hopping amplitude $t_1$ and $t_2$ are kept fixed, and we look at the edge current at the $N$-th edge of the sample. 
To start with,  at time $t = 0$, the system is initially in its ground state with parameter ($M_i, \phi_i$). At half filling the initial state of the system occupies the valence band completely. The system is now driven to a different phase by either changing a parameter abruptly or through a linear time dependent sweep from ($M_i,\phi_i$) to a new set of value ($M_f, \phi_f$). The system then unitarily evolves under the action of the new Hamiltonian, $H(M_f,\phi_f)$. 

Earlier studies \ct{caio15,alessio15} established that the Chern number of the initial ground state of the translationally invariant Haldane model remain preserved 
throughout the post quench unitary evolution for all possible quenching protocols. However,  the preservation of the bulk topological invariant (Chern number) is not reflected in the dynamics of the boundary (edge) current. Following a sudden quench, the edge current was shown to attain a new equilibrium value close to the ground state expectation value of the edge current evaluated for the final Hamiltonian. 

Motivated by this, we investigate the time evolution of the edge current for a Haldane model on a nano-ribbon geometry subjected to a slow quench, a linear time dependent sweep, from one phase to the other. 
To this end, we start with the system in the ground state of the initial Hamiltonian with parameter ($M_i, \phi_i$). Now the Semenoff mass $M(t)$ is changed linearly with time over a given interval,  keeping $\phi$ fixed, such that the final state is specified by point ($M_f, \phi_f=\phi_i$) on the phase diagram of Fig~\ref{fig1}(d). Explicitly, the quench protocol is given by $M(t) = M_i + (M_f - M_i)t/\tau$, where $1/\tau$ specifies the rate of the ramp. 

\begin{figure}[t]
	\includegraphics[width=0.8 \linewidth]{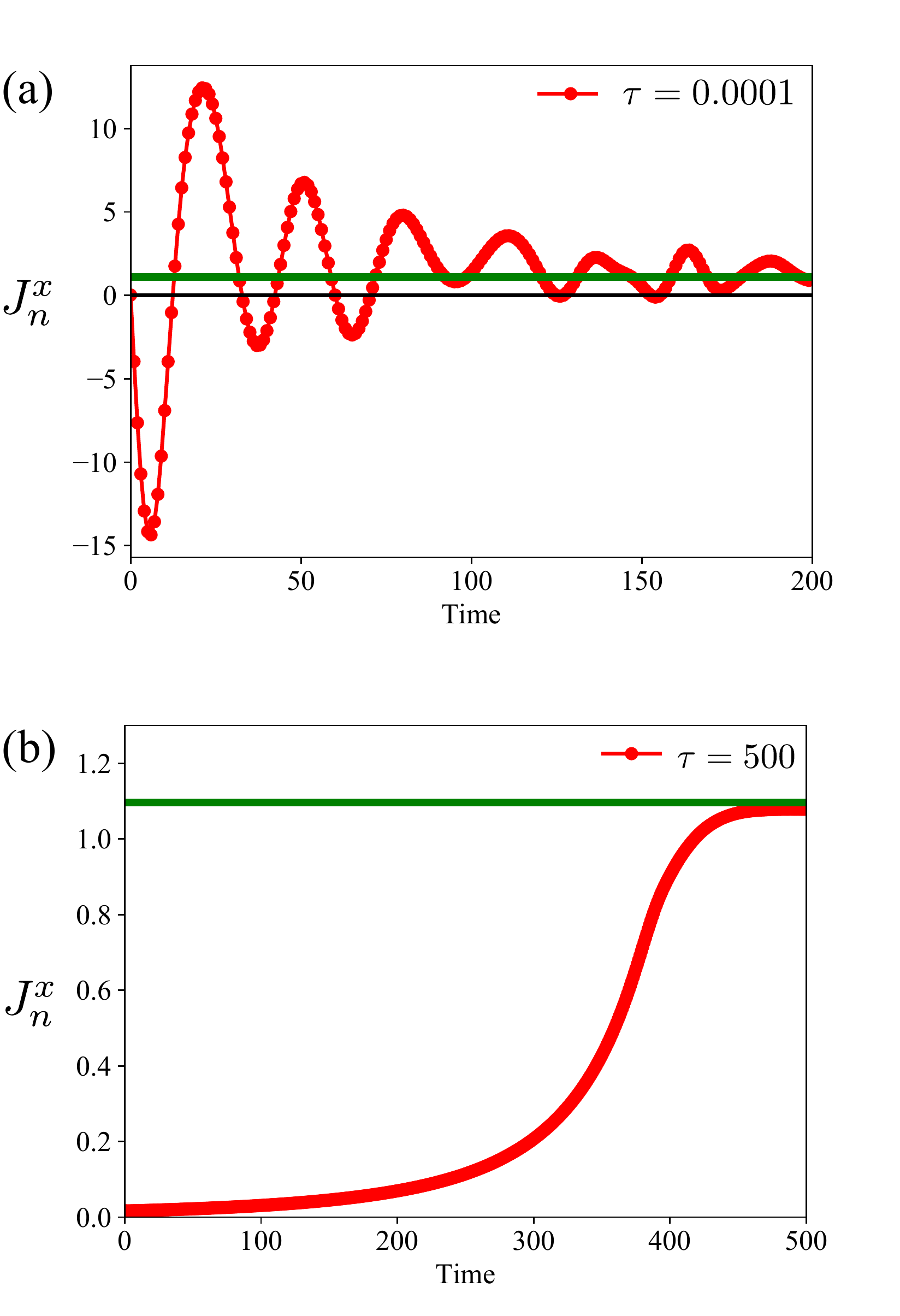}
	\caption{ Dynamics of the edge current from the non-topological phase ($M=6 , \phi = \pi/3$) to the topological phase ($M=0 , \phi = \pi/3$) following two different quenching protocols. (a) Sudden quench, when the rate of quenching is so small that the system should ideally remain in the ground state of the initial Hamiltonian and should have no current. However, the finiteness of $\tau$ leads to some excited states, which give a finite contribution to the current. 
	(b) A linear slow quench (i.e., $\tau$ large) from the trivial phase to the topological. The system maintains the instantaneous ground state of the time evolved Hamiltonian, at each instant of time, and post quench it reaches the actual ground state current of the final Hamiltonian.}\label{fig_currt}
\end{figure}

Since the problem is analytically intractable, the final edge current is numerically calculated by taking the expectation value of the current operator in Eq.~\ref{eq_curr} along the $x-$direction, with respect to the time-evolved initial state of the system obtained after solving the $2N$ coupled linear time dependent equations for every value of $k_x$ keeping $M_i,M_f$, and $\phi$ fixed throughout. 
The time evolution of the edge current, when we sweep our system ($L = 20$) from non-topological ($M = 6, \phi = \pi/3$) to topological ($M = 0, \phi = \pi/3$) phase followed by a unitary evolution, 
is shown in 
Fig.~\ref{fig_currt}. Here the edge current is shown for the $n=20$ edge, and the current at the opposite edge ($n=1$) is of the same magnitude but flows in opposite direction. 
For the particular case of $\tau \to 0 $, we are in the sudden quench regime and for $\tau\to\infty$, the quench is adiabatic. 

\subsection{Sudden quench limit}

In both panels of Fig.~\ref{fig_currt}, initially the system is in a non-topological phase and consequently the edge current always starts from its equilibrium value of zero. For the case of sudden quench, $\tau=0$, the system remains `frozen' in the initial state, i.e., the ground state of the system in the non-topological phase, and thus the current following a sudden quench is therefore zero. Nonetheless, Fig.~\ref{fig_currt} (a) shows a small but finite value of the current post quench, owing to the finiteness of $\tau$. In the case of finite but small $\tau$, the system gets excited to higher energy states of the system as well, all of which eventually do a free evolution with the final Hamiltonian. Thus the edge current is primarily governed by the  overlap between the initial ground state and the eigenstates of the final Hamiltonian, see Fig.~\ref{fig_currt}(a). The oscillations in the edge current in Fig.~\ref{fig_currt}(a) are the finite size based resurgent oscillations\cite{caio15}. 

\subsection{Adiabatic quench}
The opposite limit of the adiabatic quench, can be understood by employing the Landau-Zener argument for two mid-gap states. For no diabatic transitions (mixing of energy levels) we have $\tau \gg 1/{\Delta^2}$, where $\Delta$ is the equilibrium gap in the spectrum for the first excited state.  Now since the energy gap in our system scales inversely with $L$, we have  $\tau > L^2$ for an adiabatic evolution of the system - in which the system state follows the instantaneous ground state of the time evolved Hamiltonian at all times. Thus for a system size of $L = 20$, the adiabatic limit is achieved for a value of $\tau > 400$. Consequently, in Fig.~\ref{fig_currt}(b), we see that the current reaches a finite value infinitesimally close to the equilibrium current of the final Hamiltonian with $M=M_f$ and $\phi=\phi_f$, as expected. 
 
\begin{figure}[t]
	\includegraphics[width=0.9 \linewidth]{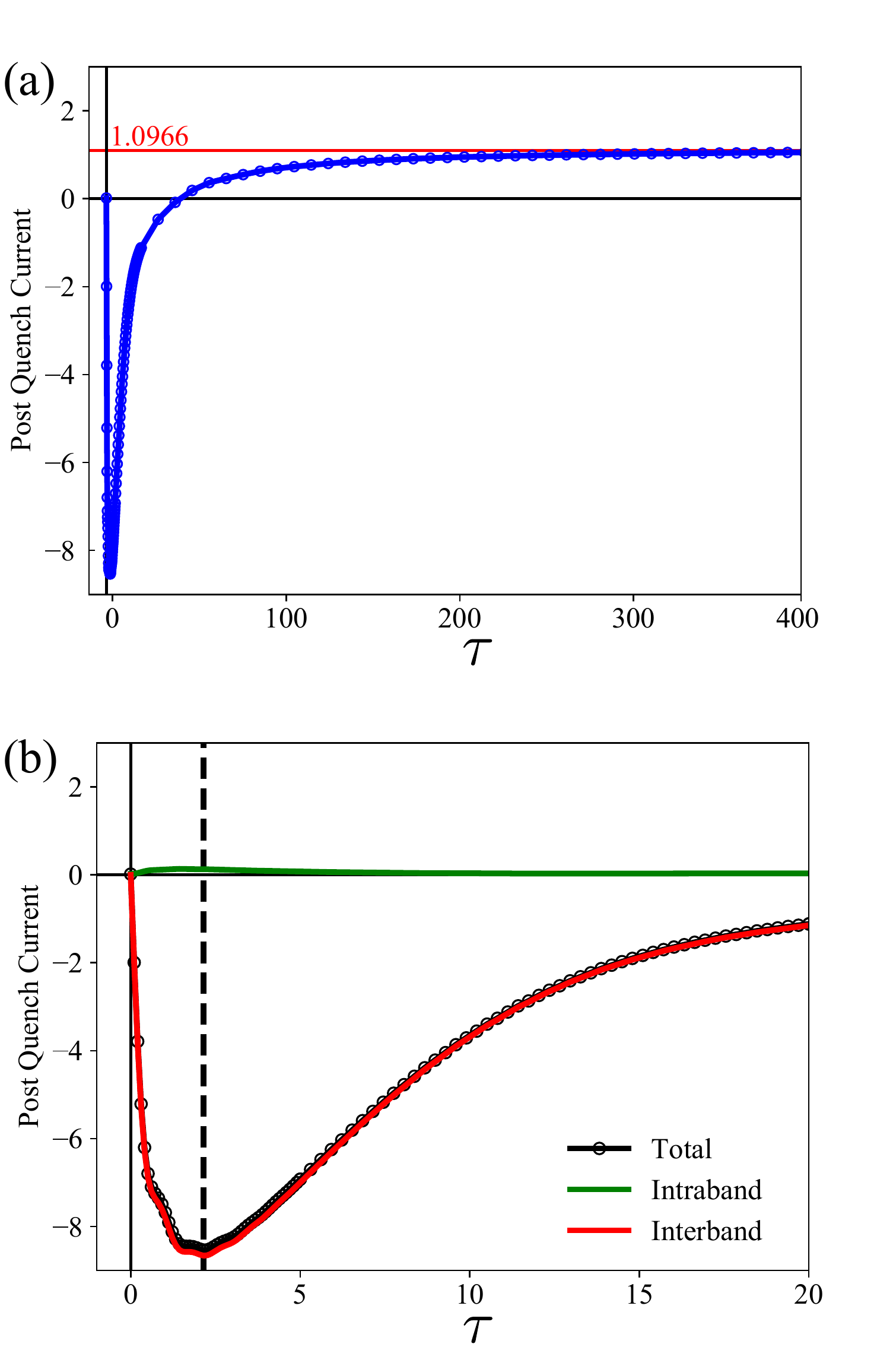}
	\caption{ (a) The variation of post quench current for a linear sweep from a non-topological  to a topological phase with different quenching rates $\tau$.  Here $\phi = \pi/3$,  $t_1=1$, and $t_2=1/3$ are held fixed while the Semenoff mass $M$ is varied in time to quench to different phases. The curve shows a generation of excess current for smaller values of $\tau$ which reaches a minima at $\tau_c=2.2$, following which it increases again to reach the equilibrium current value for the final Hamiltonian for large $\tau$. 
(b) The zoomed version of the $\tau \in (0,20)$ region of panel (a), with the intra (green) and inter (blue) band contributions [see Eq.~\eqref{eq_intra}] shown separately. Clearly excited states play a significant role, since the current is dominated by the inter band contribution. 
}\label{fig6}
\end{figure}

\begin{figure}[t]
	\includegraphics[width=1.0 \linewidth]{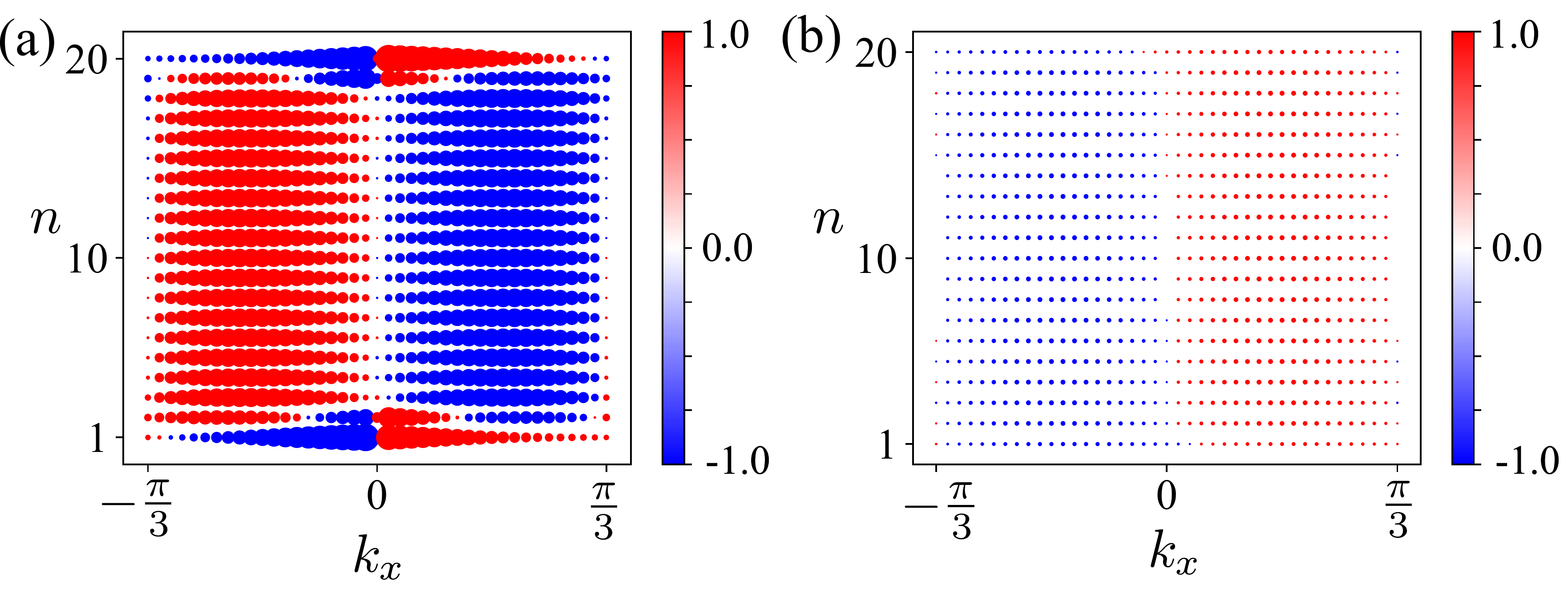}
	\caption{The spatial $n$ (see Fig.~\ref{fig_lattedge}) and the $k_x$ resolved current in the (a) topological phase and (b) the non-topological phase. The size of the dot represents the amount of current at that point and the color denotes the sign of the current (blue is for negative and red is for positive). The existence of finite edge current in the topological phase [panel (a)] and no current in the non-topological phase [panel (b)] is evident. 
	}\label{fig7}
\end{figure}

\section{Edge current dynamics following a finite time linear quench protocol} 
\label{S4}
Following the brief discussion of the sudden and the adiabatic quench case, we now turn our attention to the more interesting case of the dynamics during the intermediate times between the extreme sudden and the slow limits. To be specific, we drive the system from the non-topological ($M = 6$) to the topological ($M = 0$) phase keeping $\phi = \pi/3$ fixed, 
with different $\tau$ (varying from $0.0001-500$) and calculate the current at the $N$-th edge for $t_f =\tau$, just as the quenching stops. Naively we can expect the ramp up of the edge current from zero for $\tau \to 0$ to the finial equilibrium value for $\tau \gg L^2$ to be monotonic. However it turns out that this is not the case. 
 
The variation of the post quench current as a function of $\tau$ is shown in Fig.~\ref{fig6}(a). To start with in the topologically trivial phase, for $\tau =0$ the current is zero as expected. On the other hand for $\tau = 400~(\approx L^2$ with $L=20$), i.e., in the slow quench regime, the system always follows the instantaneous ground state of the time evolved Hamiltonian, and eventually reaches the ground state of the topological phase with $M = 0$ and $\phi = \pi/3$. Thus the edge current (calculated at $t =\tau$) also reaches its equilibrium value in the topological phase, for large $\tau \sim L^2$. 
Interestingly, the evolution of the edge current with $\tau$, is not monotonic. Starting from zero edge current in the sudden quench regime, the edge current first decreases till a critical value of $\tau = \tau_c$, and then increases again with increasing $\tau$ to finally reach its adiabatic limit equilibrium value. For the particular linear quench protocol with $M_i=6,M_f=0,\phi=\pi/3$ shown in Fig.~\ref{fig6} (a)-(b), we find that $\tau_c = 2.2$. Remarkably, the absolute value of the current at $\tau=\tau_c$ is significantly larger than the absolute value of the edge current as $\tau\to\infty$. 

The current operator can also be written as a sum of two parts, the interband and
intranband current in terms of the eigenstates of the initial Hamiltonian. We have
$\langle J^x_n \rangle = \sum_{r} \langle J^x_n \rangle_r$, where $r$ simply denotes
the index of the occupied bands, and 
\begin{eqnarray}\label{eq_intra} \nn
\langle \hat{J_n}^x\rangle_r &=& \sum_{k_x}\langle \psi^{k_x}_r (0) |
U_{k_x}^{\dagger} (t) \hat{J}_{n,k_x}^x U_{k_x}(t) | \psi^{k_x}_r(0) \rangle, \\ \nn 
&=&\sum_{p}\sum_{k_x}\langle \psi^{k_x}_r (t) 
|e_p^{k_x}\rangle\langle e_p^{k_x}|\hat{J}_{n,k_x}^x|e_p^{k_x}\rangle\langle
e_p^{k_x}| \psi^{k_x}_r(t) \rangle \nn \\ 
&+&  \sum_{p,q \neq p} \sum_{k_x}\langle \psi^{k_x}_r
(t) |e_p^{k_x}\rangle\langle
e_p^{k_x}|\hat{J}_{n,k_x}^x|e_q^{k_x}\rangle\langle e_q^{k_x}| 
\psi^{k_x}_r(t) \rangle \nn \\
&=& \sum_{p=q} \sum_{k_x}|\langle \psi^{k_x}_r(t)|e_p^{k_x}\rangle|^2 \langle
\hat{J}_{n,k_x}^x \rangle_p^{intra}\nn \\
&+& \sum_{p,q \neq p} \sum_{k_x}\langle
\psi^{k_x}_r(t)|e_p^{k_x}\rangle \langle e_q^{k_x}|\psi^{k_x}_r(t)\rangle \langle
\hat{J}_{n,k_x}^x\rangle_{pq}^{inter} 
 \end {eqnarray}
%
%
Here, $|e_p\rangle$ and $|e_q\rangle$ are the eigenstates of the initial Hamiltonian, and two parts of Eq. (\ref{eq_intra}) represent the intraband and interband contribution to the total current, respectively. As shown in Fig.~\ref{fig6}(b), 
the dominant contribution to the edge current comes from the interband contribution, with the intraband contribution being relatively small.

\subsection{Edge current reversal}
To understand the reversal in the direction of the edge current, as opposed to a monotonic rise from zero with increasing $\tau$, let us focus on the small time behaviour of the 
time evolution operator: $i\partial_t U(t)=H(t)U(t)$. For an infinitesimal increment of $\delta t/2$ in time, we have 
\begin{equation}
 U\left(\frac{\delta t}{2}\right) = U(0)-iH(0)U(0)\frac{\delta t}{2}. \label{eq_difft}\\ 
\end{equation}
Propagating to another increment of $\delta t/2$ interval, 
\begin{equation}
 U(\delta t) = U\left(\frac{\delta t}{2}\right)-iH\left(\frac{\delta t}{2}\right)U\left(\frac{\delta t}{2}\right)\frac{\delta t}{2}~. \label{eq_difff} 
\end{equation}
Since we are looking at small $\tau$ limit in vicinity of the sudden quench, we set $\tau = \delta t$ - the point at which the final current has to be calculated. 
Thus we have 
\begin{equation}
H\left(\frac{\delta t}{2}\right)= H(0) + V\left(\frac{\delta t}{2}\right), \label{eq_Ht}
\end{equation}
with $H(0)=H(M_i)$,  and 
\begin{equation}
V\left(\frac{\delta t}{2}\right)= \frac{(M_i-M_f)}{\tau} \frac{\delta t}{2}\Sigma_z
\end{equation}
where we have defined, 
\begin{equation}
\Sigma_z = \left(\begin{matrix}
 1 & 0 & 0 & 0 &\cdots \\
 0 & -1 & 0 & 0 & \cdots \\
 0 & 0 & 1 & 0 & \cdots \\
 0 & 0 & 0 & -1 & \cdots \\
 \vdots & \vdots & \vdots & \vdots & \ddots
 \end{matrix}\right).
 \end{equation}
Using Eq.~\ref{eq_difft} and Eq.~\ref{eq_Ht} in Eq.~\ref{eq_difff},  and restricting upto linear order in $\delta t$, a simplified form of the unitary operator can be obtained 
and it is given by
 \begin{equation}
U(\delta t)=\openone - i\left[H(M_i)-\frac{M_i-M_f}{4}\Sigma_z\right]\delta t.\label{eq_diffts}
\end{equation}
Now, the time evolved state under this unitary operator is given by 
 \begin{equation}
 |\psi_{p,k}(\delta t) \rangle = U(\delta t) | \psi_{p,k}(0) \rangle~, 
 \end{equation} 
 where $p$ denotes the band index and $k=k_x$. 
 Finally, the expectation value  of the edge current, to lowest order in $\delta t$,  is given by 
 \begin{eqnarray} \label{eq_currt}
 J^x(\delta t)&=&\sum_{p,k}\langle\psi_{p,k}(\delta t) |\hat{J}_k^x|\psi_{p,k}(\delta t) \rangle \nn \\ \nn
 &=&\sum_{p,k}\langle\psi_{p,k}(0) |\hat{J}_k^x|\psi_{p,k}(0) \rangle + i \left(\frac{M_i-M_f}{4}\right)\delta t \nonumber \\
 &\times &\sum_{p,k}\langle\psi_{p,k}(0) |[\hat{J}_k^x,\Sigma_z]|\psi_{p,k}(0) \rangle~.   \end{eqnarray}
Here the first term is simply the initial equilibrium current which is zero for a starting point in the non-topological phase. In Eq.~\eqref{eq_currt}, the second term brings in the effect of the time evolution for small  $\delta t = \tau$.

\begin{figure}[t]
	\includegraphics[width=0.9\linewidth]{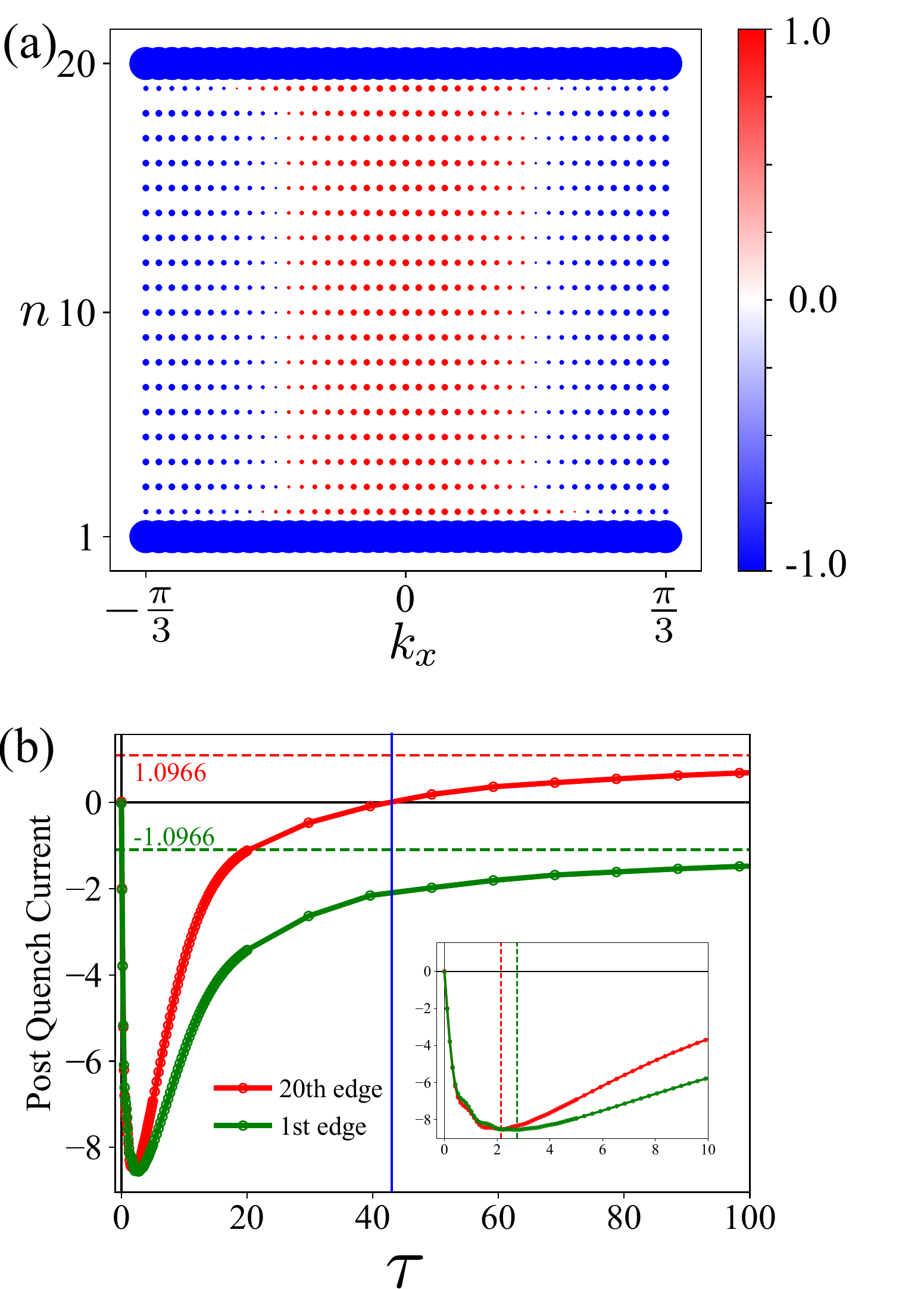}
	\caption{(a) The spatial $n$ and $k_x$ resolved contribution of the dynamical part, arising from the second term in Eq.~\eqref{eq_currt} for small $\tau$ values. The unidirectional negative contribution in both the edges, leads to the huge initial dip in current at both the edges as shown in (a). The size of the dot in (a) is proportional to the magnitude of the current. Here all the parameters are identical to that in Fig.~\ref{fig6}. (b) The variation of post quench current (calculated at $t=\tau$) with different quenching rate $\tau$ at the two edges of the system. Starting from the $\tau=0$ scenario of zero current, the current 
	 eventually goes to the respective equilibrium value for both the edges at large $\tau$, though for small $\tau$ both edges show a large dip in current and have different values of the turning point $\tau_c$ (highlighted in the inset). The solid vertical line indicates the value of $\tau$ after which the current on the two edges propagate in opposite directions. }\label{fig8}
\end{figure}
\begin{figure*}[t]
\includegraphics[width=0.99 \linewidth]{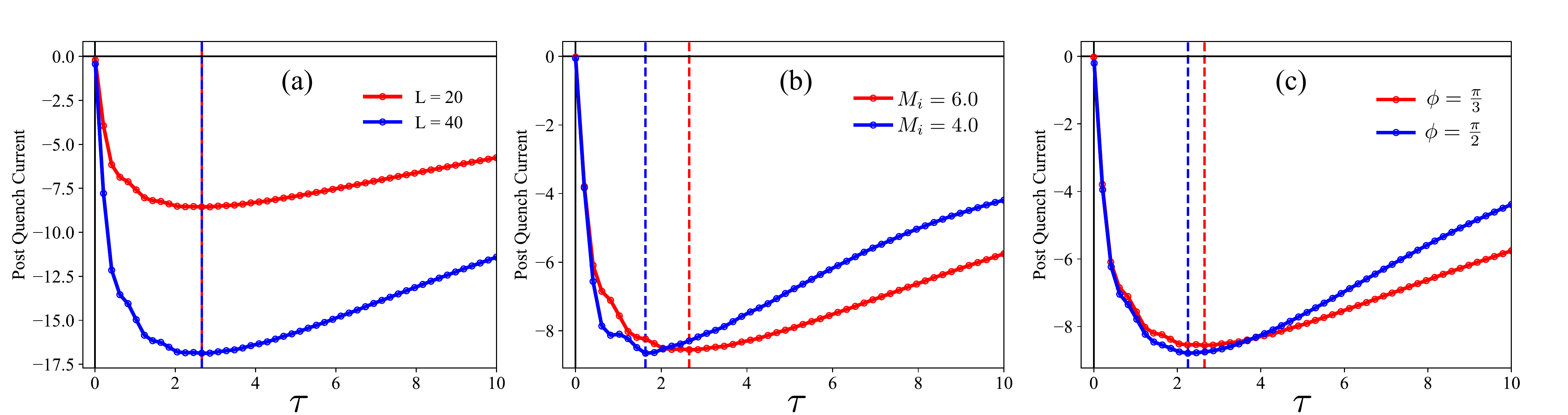}
\caption{ The dependence of $\tau_c$ on various system parameters. (a) $\tau_c$ is independent of the width of the Haldane nano-ribbon. However $\tau_c$ seems to depend on the 
parameters which determine the topological phase of the system i.e., on (b) the Semenoff mass $M$ and (c) the Haldane flux $\phi$. 
}.\label{fig9}
\end{figure*}
To understand the small time limit better, we show the spatially resolved current in the final topological phase, and the initial trivial phase in panels (a) and (b), of Fig.~\ref{fig7}, respectively. Evidently in panel (a), there is only an edge current propagating in opposite directions on the two edges in the topological phase, while there is absolutely no current to start with in the non-topological phase. The impact of the second term in Eq.~\eqref{eq_currt}, is shown in panel (a) of Fig.~\ref{fig8}. Clearly the second term in Eq.~\eqref{eq_currt}, forces a large unidirectional negative current in both the edges of the Haldane nano-ribbon at very small $\tau$. This leads to the non-monotonic behaviour of the $\tau$ dependent current, since the large negative current in both edges of Haldane nano-ribbon generated for very small small $\tau$,  has to eventually relax to the respective equilibrium values (equal in magnitude and opposite in direction) for both the edges at large $\tau$, as shown in Fig.~\ref{fig8}(b). 

One of the most interesting fact is that there is a finite value of $\tau=\tau_a$ [approximately $\tau_a \simeq 42$ in Fig.~\ref{fig8}(b)] such that for $\tau < \tau_a$  the current in both the edges is in the same direction and precisely at $\tau=\tau_a$, current in one of the edges (that carries a positive current
in the equilibrium  situation)  vanishes. This may imply a ``dynamical localization" of the current generated for $t < \tau_a$ during the ramping. On the contrary, when $\tau$ exceeds $\tau_a$, the edge current reverses sign for one of the edges implying 
that the adiabatic effect starts dominating for $\tau > \tau_a$.
While the origin of the non-monotonic behaviour of the edge current in the Haldane nano ribbon is now clear, the nature of $\tau_c$ and its dependence on various system parameters is still unknown. It turns out that $\tau_c$ does not depend on the system size at all [see Fig.~\ref{fig9}(a)], and is sensitive only to changes in the parameters deciding the topology of the phase, i.e., $M$ [see Fig.~\ref{fig9}(b)], and $\phi$ [see Fig.~\ref{fig9}(c)]. It is also different for the current on the two edges as shown in the inset of Fig.~\ref{fig8}(b).

 \section{Conclusion}\label{section4}
To summarize, we have investigated the non-equilibrium dynamics of the edge current of the semi-open Haldane model, subjected to a time dependent linear quench from the non-topological phase to the topological phase. In the sudden quench limit, the system retains its original ground state even for a quench across the phase boundary, and consequently the edge current just retains its initial value dictated by the starting phase. In case the starting point is in the non-topological phase, the edge current remains zero at all times for a sudden quench. In the opposite limit of slow quench (adiabatic limit), at each moment the system relaxes to the instantenious ground state of the time dependent Hamiltonian through out the quenching path, yielding a final edge current dictated only by the ground state current of the final Hamiltonian. 

Interestingly, we find that with increasing $\tau$, the change in the current from the initial phase current for $\tau \to 0$, to the final phase current for $\tau \to \infty$ is not monotonic. 
In the small $\tau$ limit, there is a large unidirectional current generated on both the edges of the Haldane nano ribbon. This causes the initial current to change drastically, and then with increasing $\tau$ the current on both edges relaxes to their final equilibrium value, which are equal in magnitude but opposite in direction. This leads to a non-monotonic behavior of the edge current with $\tau$. We find that the turning point for the non monotonic edge current, $\tau_c$,  is different for the two edges, does not depend on the system size, and is sensitive only to the Semenoff mass $M$, and the Haldane flux $\phi$. Furthermore, we also establish the existence of another time scale $\tau_a$ such that for $\tau > \tau_a$, the adiabatic effect starts dominating.  While most of our discussion 
is specifically for the Haldane model on a nano-ribbon, we expect similar physics to play out in other systems with topological phases and the associated edge state as well. 

\bibliography{Edge_curr_HQ}
\end{document}